\documentclass[aps,prc,nofootinbib,twocolumn,superscriptaddress,floatfix]{revtex4}
\usepackage{mathrsfs}
\usepackage{latexsym}
\usepackage{amsmath}
\usepackage{amssymb}
\usepackage{graphicx}
\usepackage{longtable}
\usepackage{multirow}
\usepackage{bbm}
\usepackage{epsfig}
\usepackage[usenames]{color}
\usepackage[colorlinks=true,citecolor=blue,linkcolor=blue]{hyperref}
\allowdisplaybreaks

\begin{document}



\title{Correlations of net baryon number and electric charge in nuclear matter}

\author{Xin-ran Yang}
\affiliation{School of Physics, Xi'an Jiaotong University, Xi'an, 710049, China}
\affiliation{ MOE Key Laboratory for Nonequilibrium Synthesis and Modulation of Condensed Matter, Xi'an Jiaotong University, Xi'an, 710049, China}

\author{Guo-yun Shao}
\email[Corresponding author: ]{gyshao@mail.xjtu.edu.cn}
\affiliation{School of Physics, Xi'an Jiaotong University, Xi'an, 710049, China}
\affiliation{ MOE Key Laboratory for Nonequilibrium Synthesis and Modulation of Condensed Matter, Xi'an Jiaotong University, Xi'an, 710049, China}

\author{Chong-long Xie}
\affiliation{School of Physics, Xi'an Jiaotong University, Xi'an, 710049, China}
\affiliation{ MOE Key Laboratory for Nonequilibrium Synthesis and Modulation of Condensed Matter, Xi'an Jiaotong University, Xi'an, 710049, China}

\author{Zhi-Peng Li}
\affiliation{School of Physics, Xi'an Jiaotong University, Xi'an, 710049, China}
\affiliation{ MOE Key Laboratory for Nonequilibrium Synthesis and Modulation of Condensed Matter, Xi'an Jiaotong University, Xi'an, 710049, China}

\begin{abstract}
\hspace{-10pt}{\bf Abstract:} We investigate the correlations between net baryon number and electric charge up to sixth order related to the interactions of nuclear matter at low temperature, and explore their relationship with the nuclear liquid-gas phase transition (LGPT) within the framework of the nonlinear Walecka model. The calculation shows that  strong correlations between the baryon number and electric charge exist in the vicinity of LGPT, and the higher order correlations are more sensitive than the lower order ones near the phase transition. However, in the high-temperature region away from the LGPT the rescaled lower order correlations are relatively larger than most of the higher order ones. Besides, some of the fifth- and sixth-order correlations possibly change the sign from negative to positive along the chemical freeze-out line with the decrease of temperature. In combination with the future experimental projects at lower collision energies, the derived results can be referred to study the phase structure of strongly interacting matter and analyze the related experimental signals.

\hspace{-10pt}{\bf Keywords:}  Correlations of conserved charges, Nuclear matter, Nuclear liquid-gas phase transition, Heavy-ion collision
 
\end{abstract}

\maketitle

 
\section{introduction}

One of the primary goals in nuclear physics is to map the phase diagram of quantum chromodynamics (QCD). It involves the chiral and deconfinement phase transtions related to  the transformation from quark-gluon plasma  to  hadronic matter~\cite{Shou24}.  The calculations from lattice QCD and hadron resonance gas~~(HRG) model indicate that a smooth crossover tranformation occurs at high temperature and small chemical potential~\cite{Aoki06, Bazavov19, Borsanyi13,Bazavov14, Bazavov17,Borsanyi14,Borsanyi20}.  
Moreover, many studies in  the effective  quark models~(e.g., Ref.\cite{Fukushima04,Ratti06,Costa10,Fu08, Sasaki12, Ferreira14, Shao2018, Schaefer10, Skokov11,Liu2018,Chen20,Zhao23,Ferreira2018,Yang24}), the Dyson-Schwinger equation approach~\cite{Gao24,Qin11,Gao16,Gao20, Fischer14,Shi14}, the functional renormalization group theory~\cite{Fu20,Rennecke17, Fu21} and machine learning~\cite{Ma23},  suggest that  a first-order chiral phase transition  undergoes at large chemical potential.

Fluctuations and correlations of conserved charges ( baryon number $B$, electric charge $Q$ and strangeness
$S$) are  sensitive observables to study the phase transition of strongly interacting matter~\cite{Stephanov}. The net proton~(proxy of net baryon) cumulants have been measured in the beam energy scan~(BES) program at the Relativistic Heavy Ion Collider (RHIC)~\cite{Aggarwal10, Adamczyk14,Chen24,Luo2017,Aboona23,Zhang23,Chen23}, which has sparked extensive study about QCD phase transition, in particular, the QCD critical endpoint~(CEP).   
More impressively, the distributions of net proton number at the center-of mass energy $\sqrt{s_{NN}}=3$ and $2.4\,$GeV are essentially different from those at $7.7\,$GeV and above, since the fluctuation distributions of net proton number are primarily dominated by the interaction among hadrons~\cite{Aboona23}.

The experimental results at $3\,$GeV and below raise the question of how the hadronic interactions affect the fluctuations of conserved charges at lower-energy regimes~\cite{Huang23, ShaoJ23, Marczenko23,Liu22}.  With the decrease of collision energy, the nuclear liquid-gas phase transition~(LGPT) is possibly involved~\cite{He23,Shao20,Chomaz04,Pochodzalla95,Borderie01,Botvina95,Agostino99,Srivastava02,Elliott02,Xu2023,Ma1999,Deng2022,Hempel13, Mukherjee17,Xu13,Savchuk20,Wang20}. 
In Ref.~\cite{Poberezhnyuk19,Vovchenko17,Poberezhnyuk21}, a van der Waals model was used to study the high-order distributions of net baryon number in both the pure and mixed phases of the LGPT. In Ref.~\cite{Marczenko23},  the second-order susceptibility of net baryon number for positive- and negative-parity nucleons was examined near the chiral and nuclear liquid-gas phase transitions using the double parity  model, in which both the chiral phase transition and nuclear LGPT are effectively included. In Ref.~\cite{Shao20,Xu2023}, the net baryon kurtosis and skewness were considered  in the non-linear Walecka model to analyze the experimental signals at lower collision energies. The hyperskewness and hyperkurtosis of net baryon number were further calculated recently  to explore the relation between nuclear LGPT and experimental observables~\cite{Yang2024}.

Since the interactions among hadrons dominate the density fluctuations at lower energy regimes~(below $3\,$GeV),  the BES program at collision energies lower than $7.7\,$GeV will provide more information about the phase structure of strongly interacting matter. The relevant experiments are also in plan at High Intensity heavy-ion Accelerator Facility (HIAF). Meanwhile, the HADES collaboration at GSI Helmholtzzentrum für Schwerionenforschung planned to measure higher-order net proton and net charge fluctuations in central Au + Au reactions at collision energies ranging from $0.2A$ to $1.0A\,$GeV to probe the LGPT region~\cite{Bluhm20}. These experiments are significant  for  investigating  the nuclear liquid-gas and chiral phase transitions through the density fluctuations.  

Besides the fluctuations of conserved charges, the correlations of different conserved charges can also provide important information to explore the phase transition. The correlations of conserved charges or the off-diagonal susceptibilities have been calculated to study the chiral and deconfinement phase transitions at high temperature in lattice QCD and some effective quark models~(e.g., \cite{Bellwied15, Wen2021, Ding2015, Fodor18, Bellwied20, Fu10, Bhattacharyya11}).  However, the correlations of net baryon number and electric charge in nuclear matter and their relationship with nuclear LGPT are still absent, which are useful in diagnosing the phase diagram of strongly interacting matter at low temperature.   
In this study, we will explore the correlations between net baryon number and electric charge up to sixth order in nuclear matter using the nonlinear Walecka model. Some characteristic behaviors of correlations evoked by the nucleon-nucleon interaction near and far away from the nuclear LGPT are obtained. These results will help analyze the chiral phase transition, nuclear LGPT and the  related experimental signals in the future.

The paper is organized as follows. In Sec.~II, we introduce the formulas to describe correlations of conserved charges and the nonlinear Walecka model. In Sec.~III, we illustrate the numerical results of  correlations of net baryon number and electric charge. A summary is finally given in Sec. IV.

\section{ Correlations of conserved charges and the nonlinear Walecka model}

The fluctuations and correlations of conserved charges are related to the equation of state of a thermodynamic system.  In the grand-canonical ensemble of strongly interacting matter the pressure is the logarithm of partition function \cite{Karsch15}:
\begin{equation}\label{}
P=\frac{T}{V}\ln{Z(V,T,\mu_B, \mu_Q, \mu_S)},
\end{equation}  
where $\mu_B, \mu_Q, \mu_S$ are the chemical potentials of conserved charges, i.e., the baryon number, electric charge and strangeness in strong interaction, respectively. The generalized susceptibilities can be derived by taking the partial derivatives of the pressure with respect to the corresponding chemical potentials \cite{Luo2017}
\begin{equation}\label{}
\chi^{BQS}_{ijk}=\frac{\partial^{i+j+k}[P/T^4]}{\partial(\mu_B/T)^i \partial(\mu_Q/T)^j \partial(\mu_S/T)^k}.
\end{equation}  

In experiments, the cumulants of multiplicity distributions of the conserved charges are usually measured. They are related to the generalized susceptibilities by
 \begin{equation}\label{Crho}
\!C^{BQS}_{ijk}\!=\!\frac{\partial^{i\!+\!j\!+\,k}\ln[Z(V,T,\mu_B, \mu_Q, \mu_S)]}{\partial(\mu_B/T)^i \partial(\mu_Q/T)^j \partial(\mu_S/T)^k}\!=\!V\!T^3\chi^{BQS}_{ijk}\, .
\end{equation} 
To eliminate the volume dependence in heavy-ion collision experiments, observables are usually constructed by the ratios of cumulants, and then can be compared with the theoretical calculations of the generalized susceptibilities with
 \begin{equation}\label{}
\frac{C^{BQS}_{ijk}}{C^{BQS}_{lmn}} =\frac{\chi^{BQS}_{ijk}}{\chi^{BQS}_{lmn}}.
 \end{equation} 

In this research the nonlinear Walecka model is taken to calculate the correlations of net baryon number and electric charge in nuclear matter at low temperature. This model is generally used to describe the properties of finite nuclei and the equation of state of nuclear matter.  The approximate equivalence of this model to the hadron resonance gas model at low temperature and small density was also indicated in Ref.~\cite{Fukushima15}. This model was recently taken to explore the fluctuations of net baryon number in nuclear matter, e.g., the kurtosis and skewness in Ref.~\cite{Shao20, Xu2023}, and the hyperskewness and hyperkurtosis~\cite{Yang2024}.

The Lagrangian density for the nucleons-meson system in the nonlinear Walecka model \cite{Glendenning97, He23} is 
\begin{eqnarray}\label{lagrangian}
	\cal{L}\!&\!=\!&\sum_N\bar{\psi}_N\!\big[i\gamma_{\mu}\partial^{\mu}\!-\!(\!m_N
	\!-\! g_{\sigma }\sigma)\!
	\! -\!g_{\omega }\gamma_{\mu}\omega^{\mu} \!-\! g_{\rho }\!\gamma_{\mu}\boldsymbol\tau_{}\!\cdot\!\boldsymbol
	\rho^{\mu} \big]\!\psi_N       \nonumber\\
	& &    +\frac{1}{2}\left(\partial_{\mu}\sigma\partial^
	{\mu}\sigma-m_{\sigma}^{2}\sigma^{2}\right)\!-\! \frac{1}{3} bm_N\,(g_{\sigma} \sigma)^3-\frac{1}{4} c\,
	(g_{\sigma} \sigma)^4
	\nonumber\\
	& &+\frac{1}{2}m^{2}_{\omega} \omega_{\mu}\omega^{\mu}
	-\frac{1}{4}\omega_{\mu\nu}\omega^{\mu\nu}  \nonumber \\
	& &    +\frac{1}{2}m^{2}_{\rho}\boldsymbol\rho_{\mu} \! \cdot \! \boldsymbol
	\rho^{\mu}   \!- \! \frac{1}{4}\boldsymbol\rho_{\mu\nu} \! \cdot \! \boldsymbol\rho^{\mu\nu} ,
\end{eqnarray}
where $
\omega_{\mu\nu}= \partial_\mu \omega_\nu - \partial_\nu
\omega_\mu$, $ \rho_{\mu\nu} =\partial_\mu
\boldsymbol\rho_\nu -\partial_\nu \boldsymbol\rho_\mu$ and $m_N$ is the nucleon mass in vacuum. The interactions between
nucleons are mediated by $\sigma,\,\omega,\,\rho$ mesons..

The thermodynamic  potential can be derived in the mean-field approximation 
\begin{eqnarray}
 \Omega=\!&\!&\!-\!\beta^{-1} \sum_{N} 2\! \int \frac{d^{3} \boldsymbol{k}}{(2 \pi)^{3}}\!\bigg[\ln \!\left(1+e^{-\beta\left(E_{N}^{*}(k)-\mu_{N}^{*}\!\right)}\right)\! \nonumber \\ 
 &&+\!\ln\! \left(1\!+\!e^{-\beta\left(E_{N}^{*}(k)+\mu_{N}^{*}\right)}\right)\!\bigg]\! +\!\frac{1}{2} m_{\sigma}^{2} \sigma^{2}\!+\!\frac{1}{3} b m_{N}\left(g_{\sigma} \sigma\right)^{3} \nonumber \\
 &&+\frac{1}{4} c\left(g_{\sigma} \sigma\right)^{4}-\frac{1}{2} m_{\omega}^{2} \omega_{}^{2}-\frac{1}{2} m_{\rho}^{2} \rho_{3}^{2}  ,
  \end{eqnarray}
where $\beta=1/T$, $E_{N}^{*}=\sqrt{k^{2}+m_{N}^{*2}}$,  and  $\rho_{3}$ is the third component of $\rho$ meson field. The effective nucleon mass $m^*_N=m_N- g_\sigma \sigma$ and the effective chemical potential $\mu_{N}^{*}$ is defined as  $\mu_{N}^{*}=\mu_{N}-g_{\omega} \omega_{}-\tau_{3 N} g_{\rho} \rho_{3}$ ($\tau_{3 N}=1/2$ for proton, $-1/2$ for neutron).

By minimizing the thermodynamical potential  
\begin{equation}
\frac{\partial \Omega}{\partial \sigma}=\frac{\partial \Omega}{\partial \omega_{}}=\frac{\partial \Omega}{\partial \rho_{3}}=0,
\end{equation}
the meson field equations can be derived as
\begin{equation}\label{sigma}
g_{\sigma} \sigma\!=\!\left(\frac{g_{\sigma}}{m_{\sigma}}\right)^{2}\!\left[\rho_{p}^{s}+\rho_{n}^{s}-b m_{N}\left(g_{\sigma} \sigma\right)^{2}-c\left(g_{\sigma} \sigma\right)^{3}\right],
\end{equation}
\begin{equation}
g_{\omega} \omega=\left(\frac{g_{\omega}}{m_{\omega}}\right)^{2}\left(\rho_{p}+\rho_{n}\right),
\end{equation}
\begin{equation}\label{rho}
g_{\rho} \rho_{3}=\frac{1}{2}\left(\frac{g_{\rho}}{m_{\rho}}\right)^{2} \left(\rho_{p}-\rho_{n}\right).
\end{equation}
In Eqs.(\ref{sigma})-(\ref{rho}), the nucleon number density
\begin{equation}
\rho_i=2 \int \frac{d^{3} \boldsymbol{k}}{(2 \pi)^{3}} [f\left(E_{i}^{*}-\mu_{i}^{*}\right)-\bar{f}\left(E_{i}^{*}+\mu_{i}^{*}\right) ],
\end{equation}
and the scalar density
\begin{equation}
\rho_{i}^{s}=2\int \frac{d^{3}\boldsymbol{k}}{(2 \pi)^{3}} \frac{m_{i}^{*}}{E_{i}^{*}}[f\left(E_{i}^{*}-\mu_{i}^{*}\right)+\bar{f}\left(E_{i}^{*}+\mu_{i}^{*}\right)],
\end{equation}
where $f(E_{i}^{*}-\mu_{i}^{*})$ and $\bar{ f} (E_{i}^{*}+\mu_{i}^{*})$ are the  fermion and antifermion distribution functions with
\begin{equation}
    f(E_i^*-\mu_i^*)=\frac1{1+\exp\left\{\left[E_i^*-\mu_i^*\right]/T\right\}} ,
\end{equation}
and
\begin{equation}
    f(E_i^*+\mu_i^*)=\frac1{1+\exp\left\{\left[E_i^*+\mu_i^*\right]/T\right\}} .
\end{equation}

The meson field equations can be solved for a given temperature and chemical potential (or baryon number density). 
The model parameters, $g_\sigma, g_\omega, g_\rho, b$ and $ c$, are listed in Table.~\ref{tab:1}. They are fitted with the compression modulus $K=240\,$MeV, the symmetric energy  $a_{sym}=31.3\,$MeV, 
the effective nucleon mass $m^*_N=m_N- g_\sigma \sigma=0.75m_N$ and the
 binding energy $B/A=-16.0\,$MeV at nuclear saturation density with $\rho_0=0.16\, fm^{-3}$. 
 \begin{table}[ht]
	\centering
	\caption{Parameters in the nonlinear Walecka model}
	\label{tab:1}
	\begin{tabular*}{\columnwidth}{@{\extracolsep{\fill}}lllll@{}}
	\hline
	\multicolumn{1}{@{}l}{$\left({g_\sigma }/{m_\sigma } \right) ^2$/fm$^2$} & $\left({g_\omega  }/{m_\omega  } \right) ^2$/fm$^2$ & $\left({g_\rho }/{m_\rho } \right) ^2$/fm$^2$ & $b$ & $c$ \\
	\hline
	 $ 10.329$                   & 5.423        &  0.95      & 0.00692     &  -0.0048        \\ 
	 \hline
	\end{tabular*}
	\end{table}

\section{numerical results and discussions}
In this section, we present the numerical results of the correlations between net baryon number and electric charge in  the non-linear Walecka model.  To simulate the physical conditions in the BES program at RHIC STAR, the isospin asymmetric nuclear matter is considered in the calculation with the constraint of $\rho_Q/\rho_B=0.4$. In the present Walecka model,  strange baryons are not included,  thus the strangeness condition of  $\rho_S=0$ is automatically satisfied. Note that $\rho_Q/\rho_B = 0.4$ might be slightly deviated due to isospin dynamics. We will detailedly explore the influence of different isospin asymmetries on the fluctuations and correlations of conserved charges in a separate study.

\begin{figure}[htbp]
	\begin{center}
		\includegraphics[scale=0.32]{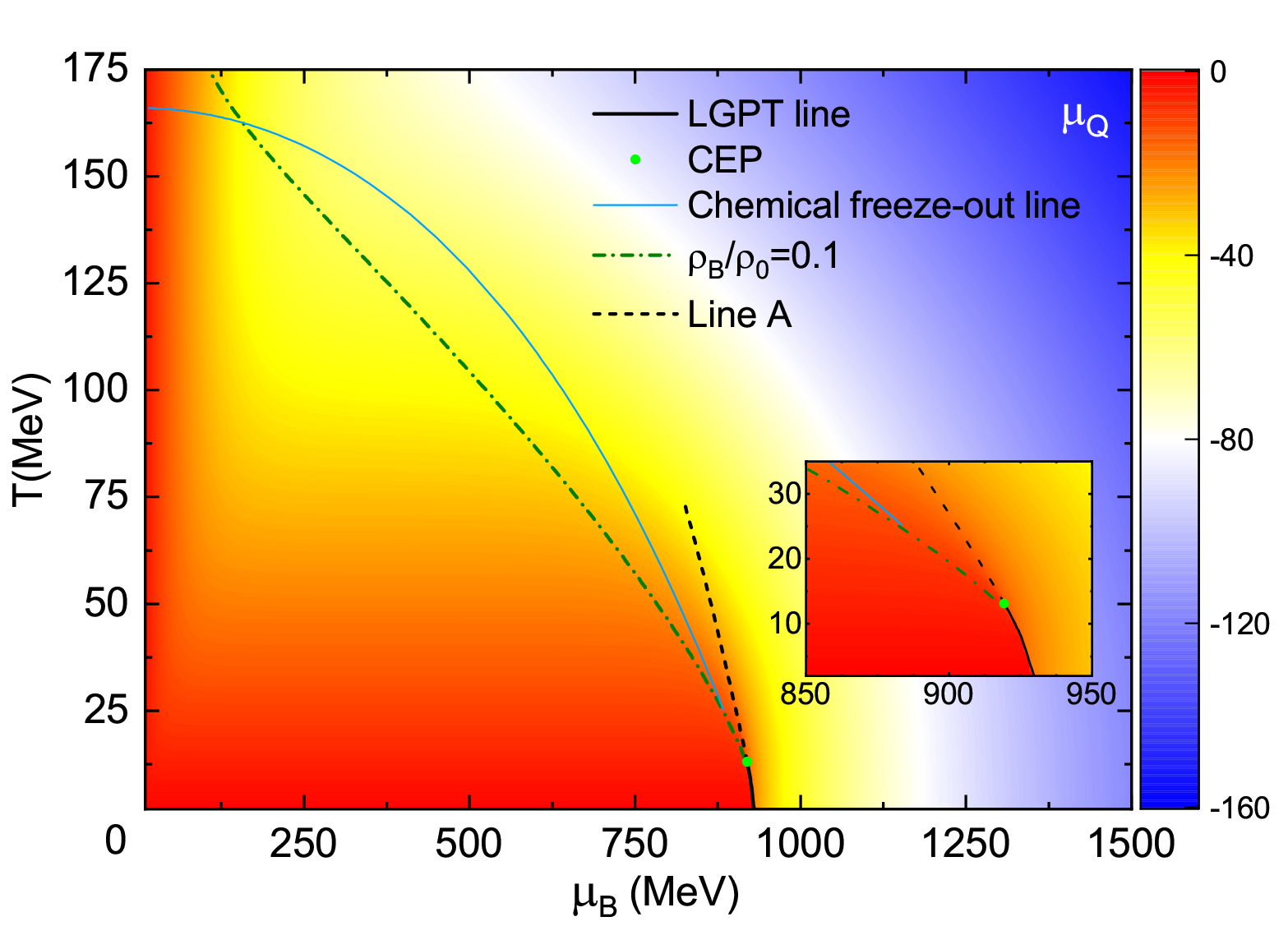}
		\caption{\label{fig:1} 
			Contour of $\mu_Q$ in the  $T-\mu_B$ plane derived in the nonlinear Walecka model with the constraint of $\rho_Q/\rho_B=0.4$. The solid line is the liquid-gas transition line with a CEP locating at $T=13\,$MeV and $\mu_B=919\,$MeV. The blue line is the chemical freeze-out line fitted in Ref.~\cite{Cleymans06}. The dash-dotted line corresponds to the temperature and chemical potential for $\rho_B=0.1\rho_0$. ``Line A" is derived with $\partial \sigma / \partial \mu_B$ taking the maximum value for each given temperature. }
	\end{center}
\end{figure}

The correlations between baryon number and electric charge are related to the baryon chemical potential $\mu_B$ and isospin chemical potential $\mu_Q$~($\mu_Q=\mu_p-\mu_n$). 
To demonstrate the value of $\mu_Q$ as a function of temperature and baryon chemical potential,
we first plot in Fig.~\ref{fig:1} the contour map of $\mu_Q$ in the  $T-\mu_B$ plane derived under the constraint of $\rho_Q/\rho_B=0.4$. The corresponding liquid-gas phase transition line with a CEP locating at $T=13\,$MeV and $\mu_B=919\,$MeV is also plotted in this figure. To compare with the chiral crossover phase transition of quarks, the dashed line labeled as ``Line A" in Fig.~\ref{fig:1} is derived with the condition that  $\partial \sigma / \partial \mu_B$ takes the maximum value for each given temperature. This line plays a role analogous in a certain degree to the chiral crossover transformation, although it is not a true phase transition in nuclear matter. It indicates the location where the dynamical nucleon mass changes most quickly with the increase of chemical potential.
The reason for this is to emphasize that both the $\sigma $ field in nuclear matter and quark condensate in quark matter are associated with the dynamical mass of fermions and, 
therefore the rapid change of mass might have the universal effect on fluctuation distributions of conserved charges.
As pointed out in our previous studies~\cite{He2023,Shao20, Yang2024}, the location of line A helps understand the behaviors of interaction measurement~(trace anomaly), the fluctuations of conserved charges near the phase transition~\cite{Shao20, Yang2024}.

One can also define “Line A” by the maximum point of $\partial\omega/\partial\mu_B$ or $\partial n_B/\partial\mu_B$, since the density can be taken as the order parameter for liquid-gas phase transition. Under this definition, the result obtained in quark model does not correspond to the chiral crossover phase transition. This is not the purpose of this study. Our aim is to indicate some common properties related to dynamical fermion mass near the critical region of a first-order phase transition.
On the other hand, the calculation indicates that the curves (“Line A”) under the two definitions coincide near the critical region, and the two curves gradually deviate at higher temperatures away from the critical region.

For the convenience of subsequent discussion of experimental observables, we also plot in Fig.~\ref{fig:1} the  chemical freeze-out line fitted with experimental data at high energies~\cite{Cleymans06}, which can be described with 
\begin{equation}\label{fl}
	T\left(\mu_B\right)=a-b \mu_B^2-c \mu_B^4,
\end{equation}
where $a=0.166\, \mathrm{GeV}, b=0.139 \, \mathrm{GeV}^{-1}$ and $c=0.053 \, \mathrm{GeV}^{-3}$. 
We should remind that the trajectories of the present relativistic heavy-ion collisions do not pass through the $T_C$ of nuclear LGPT. It is still not known how far the realistic chemical freeze-out line is from the critical region in future experiments. However, similar to the chiral phase transition of quarks, the existence of nuclear LGPT affects the fluctuation and correlation of net baryon and electric charge number in the region not very adjacent to the critical end point in intermediate-energy heavy-ion collision experiments. The numerical results on the parameterized chemical freeze-out line in this study can be taken as a reference. The realistic chemical freeze-out condition at intermediate and low energies will be extracted in future heavy-ion collision experiments.  When analyzing the experimental data the contribution from LGPT needs to be considered.

Fig.~\ref{fig:1} shows that  the value of  $\left| \mu_Q \right|$ is smaller than $40\,$MeV in the area covered in red. In this region the baryon number density is very small, which can be seen roughly from the temperature and chemical potential curve for $\rho_B=0.1\rho_0$~(dash-dotted line).   The value of $\left| \mu_Q \right|$ increases with the rising baryon density~(corresponding to larger chemical potential).  This trend of   $\left| \mu_Q \right|$ is  clearly illustrated in Fig.~1. Along the chemical freeze-out line~(solid blue line), one can see how  $\mu_Q$ changes at freeze-out with the decrease of temperature or collision energy.
\begin{figure}[htbp]
	\begin{center}
		\includegraphics[scale=0.3]{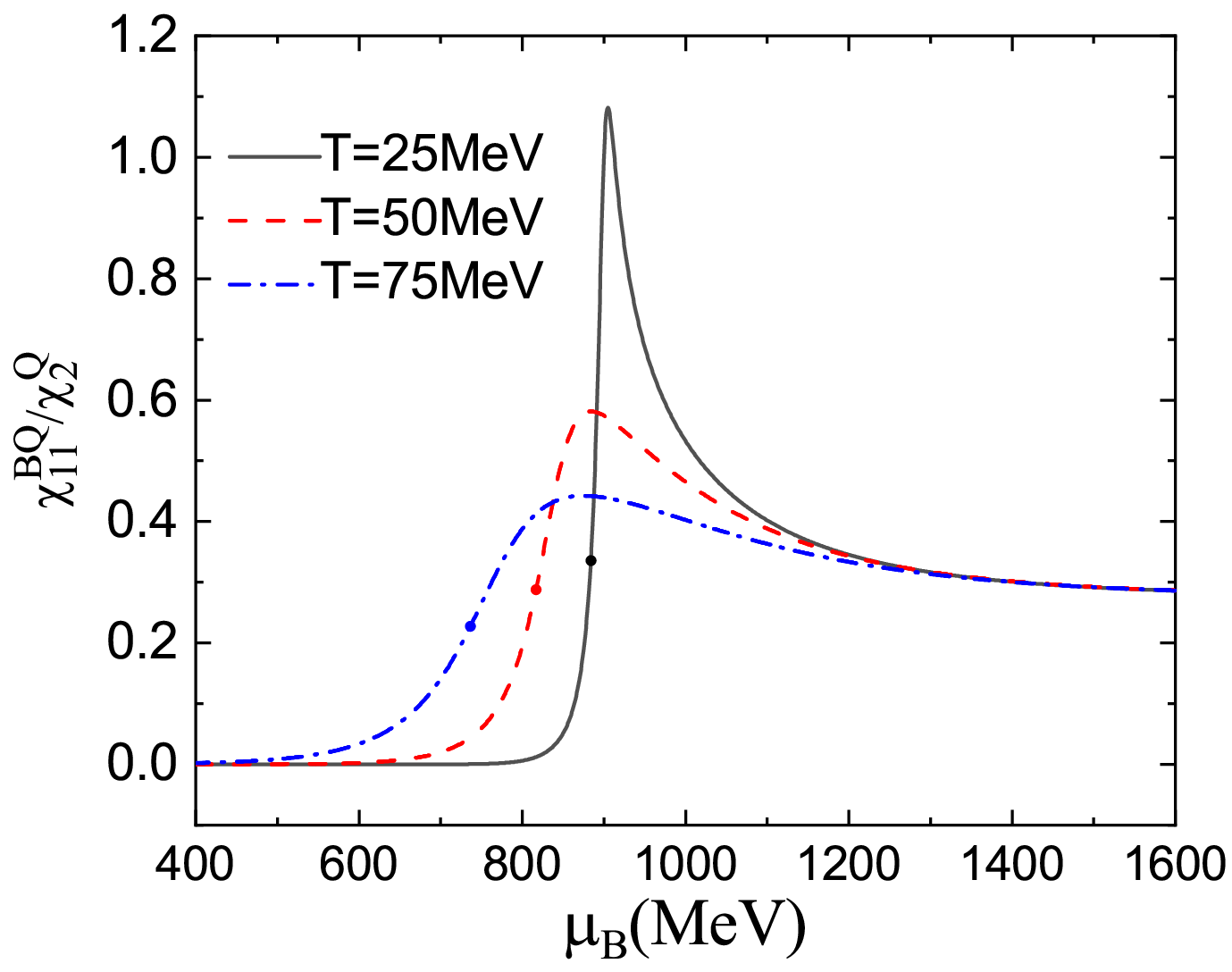}
	\end{center}
	\caption{\label{fig:2} Second order correlation between baryon number and electric charge as a function of chemical potential for different temperatures. The solid dots demonstrate the values on the chemical freeze-out line given in Fig.~\ref{fig:1}.
	}
\end{figure}

We demonstrate in Fig.~\ref{fig:2} the second order correlation between baryon number and electric charge, $\chi_{11}^{BQ}/\chi_2^Q$, as functions of baryon chemical potential for $T=75, 50, 25\,$MeV, respectively. To derive the physical quantity comparable with future experiments the correlated susceptibility is divided by $\chi_2^Q$, which eliminates the volume dependence. For each temperature, the rescaled second-order correlation $\chi_{11}^{BQ}/\chi_2^Q$ in Fig.~\ref{fig:2} displays a nonmonotonic behavior with a peak structure at a certain chemical potential. The values of these peaks increase with the decline of temperature, which indicate the correlation between baryon number and electric charge is enhanced near the phase transition region. The solid dots in Fig.~\ref{fig:2} demonstrate the values at chemical freeze-out described by Eq.~(\ref{fl}), which illustrate that  the value of $\chi_{11}^{BQ}/\chi_2^Q$ increases along the  freeze-out line when moving from the high-temperature side to the critical region.

\begin{figure}[htbp]
	\begin{center}
		\includegraphics[scale=0.3]{B1Q1.eps}
		\includegraphics[scale=0.3]{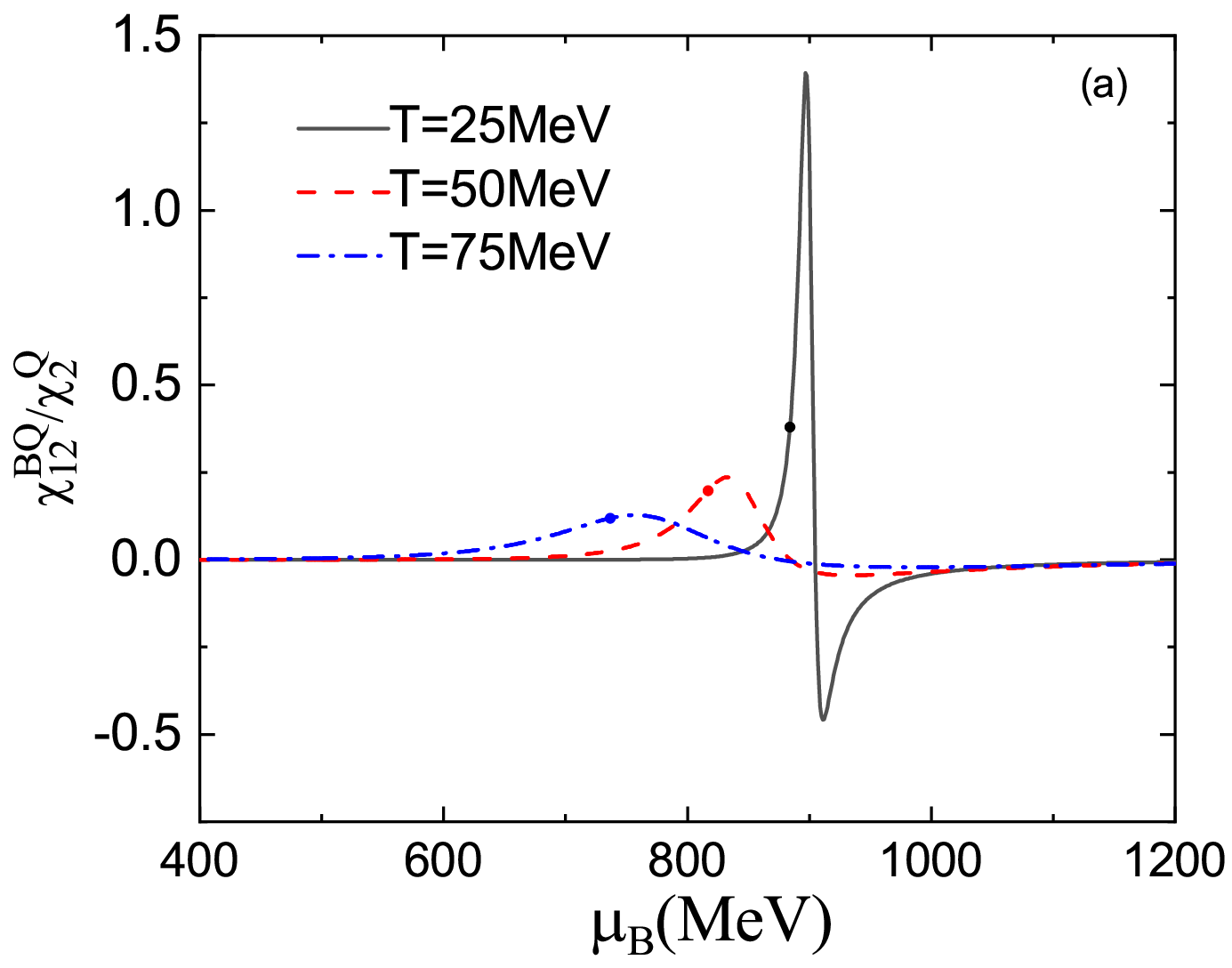}
		\includegraphics[scale=0.3]{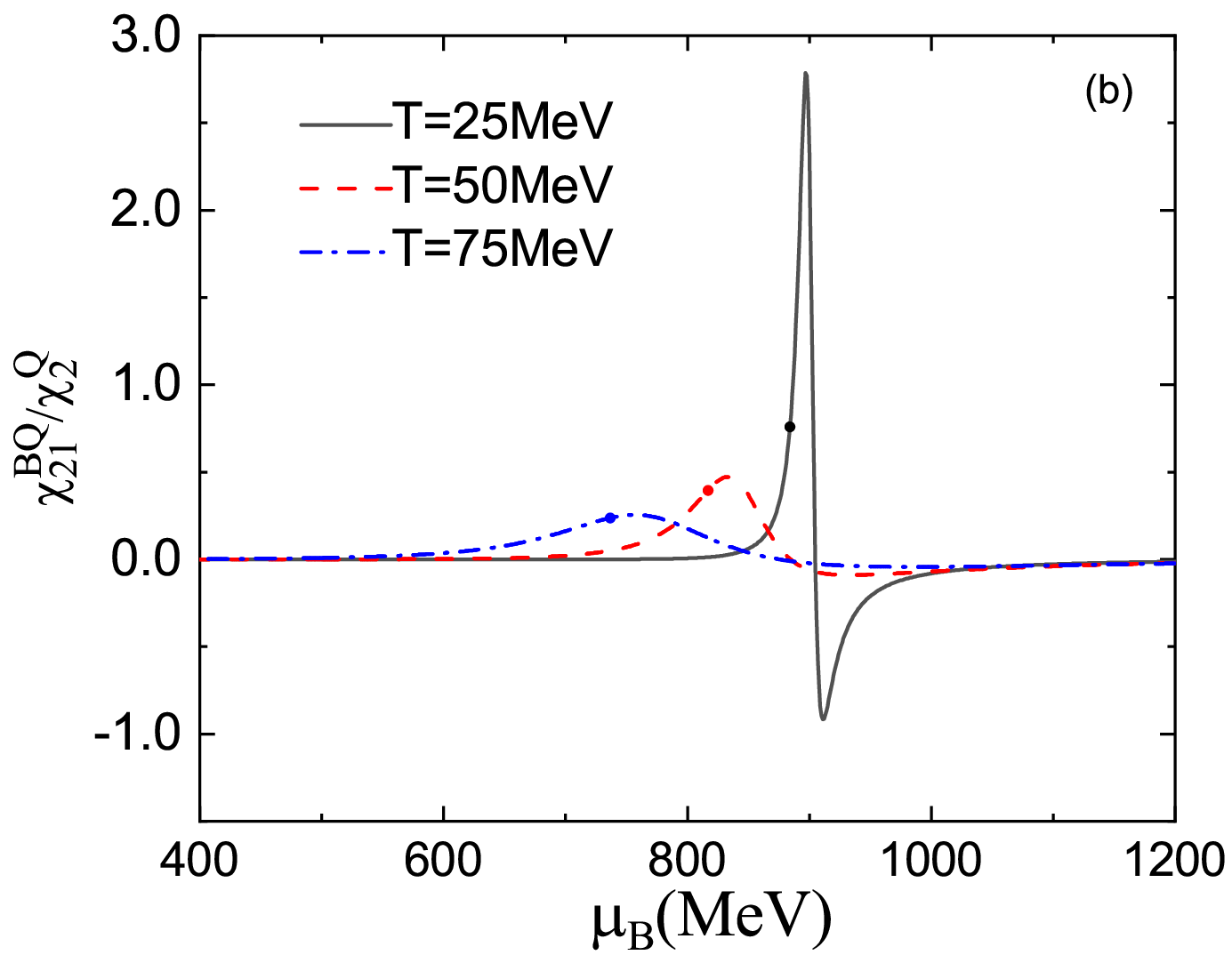}
	\end{center}
	\caption{\label{fig:3} Third order correlations between baryon number and electric charge as functions of chemical potential at different temperatures. The solid dots demonstrate the values on the chemical freeze-out line plotted in Fig.~\ref{fig:1}.
	}
\end{figure}

Fig.~\ref{fig:3} shows the third order correlations,  $\chi_{12}^{BQ}/\chi_2^Q$ and $\chi_{21}^{BQ}/\chi_2^Q$, as functions of chemical potential for several temperatures. Compared with the $\chi_{12}^{BQ}/\chi_2^Q$,  the fluctuation of  $\chi_{21}^{BQ}/\chi_2^Q$ is relatively larger at the same temperature. The solid dots at chemical freeze-out line present the same trend. This means the measurement of $\chi_{21}^{BQ}/\chi_2^Q$ is more sensitive than $\chi_{12}^{BQ}/\chi_2^Q$ in heavy-ion collision experiments. Fig.~\ref{fig:3} also indicates that with the decrease of temperature, the correlations between baryon number and electric charge intensify.
An evident oscillations  of $\chi_{12}^{BQ}/\chi_2^Q$ and $\chi_{21}^{BQ}/\chi_2^Q$ appear for $T=25\,$MeV, accompanied by the alternating positives and negatives. With the decrease of temperature, the divergent behavior appears at the CEP of LGPT. These features can be used to look for the signal
 of phase transition in experiments.

\begin{figure}[htbp]
	\begin{center}
		\includegraphics[scale=0.3]{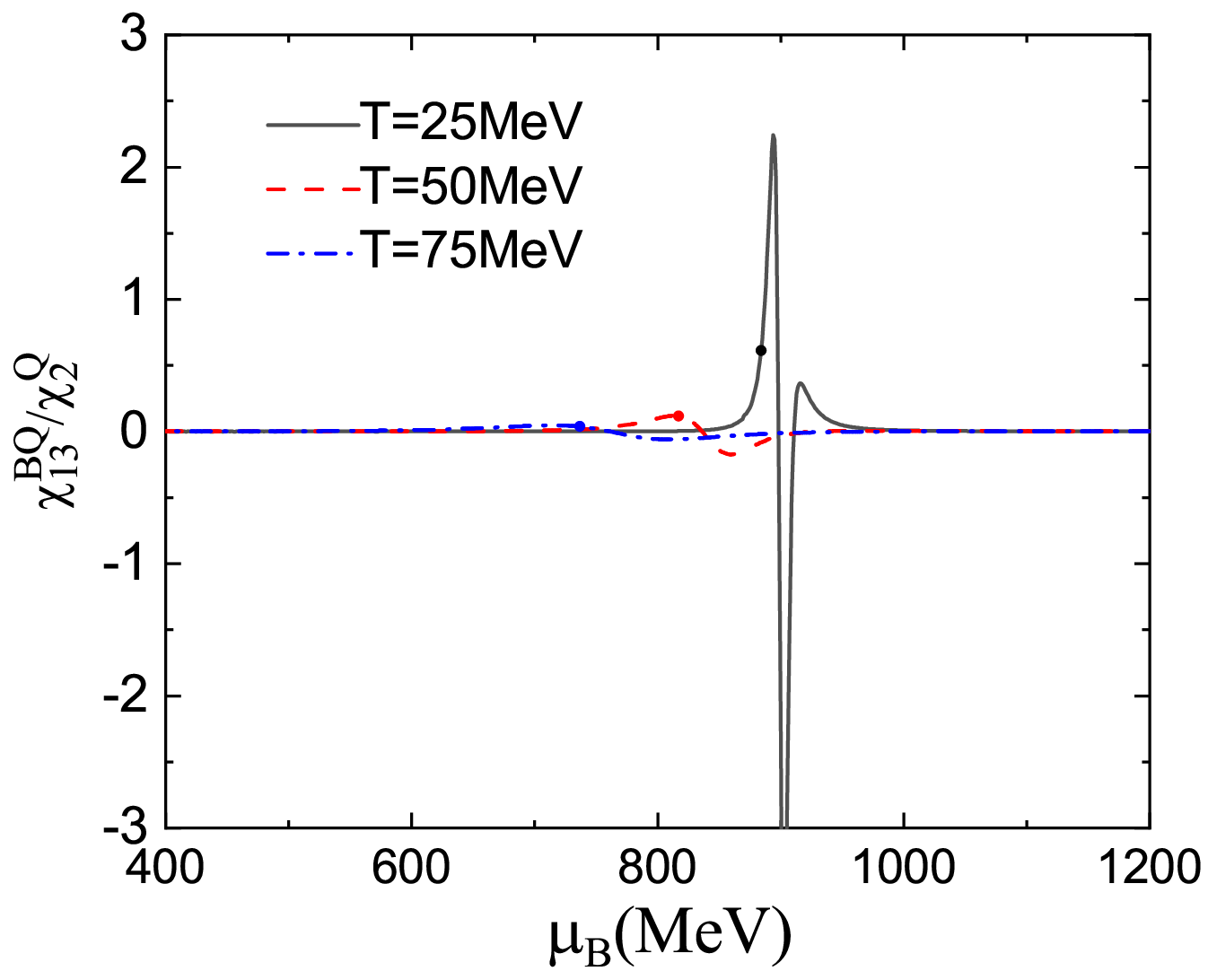}
		\includegraphics[scale=0.3]{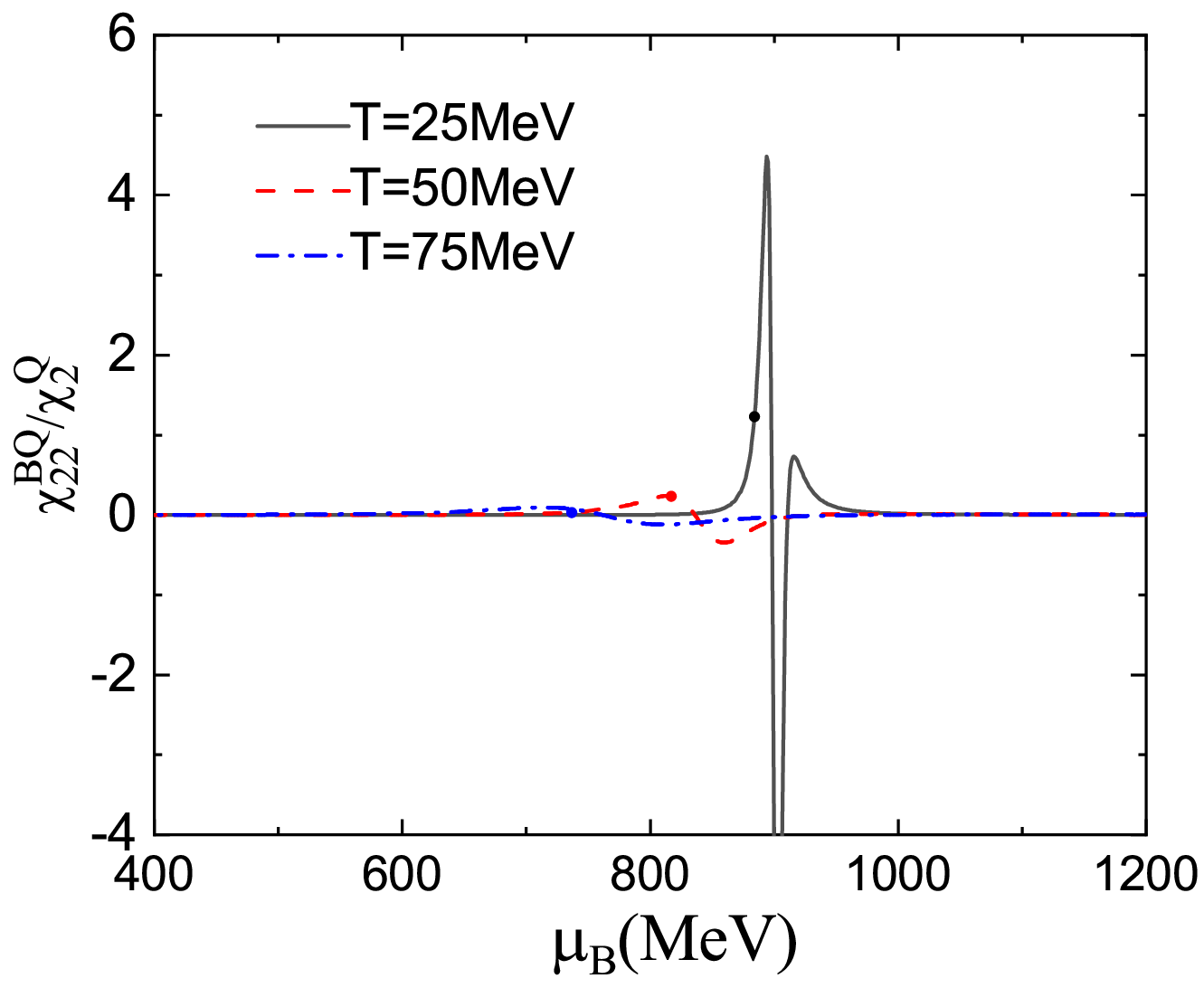}
		\includegraphics[scale=0.3]{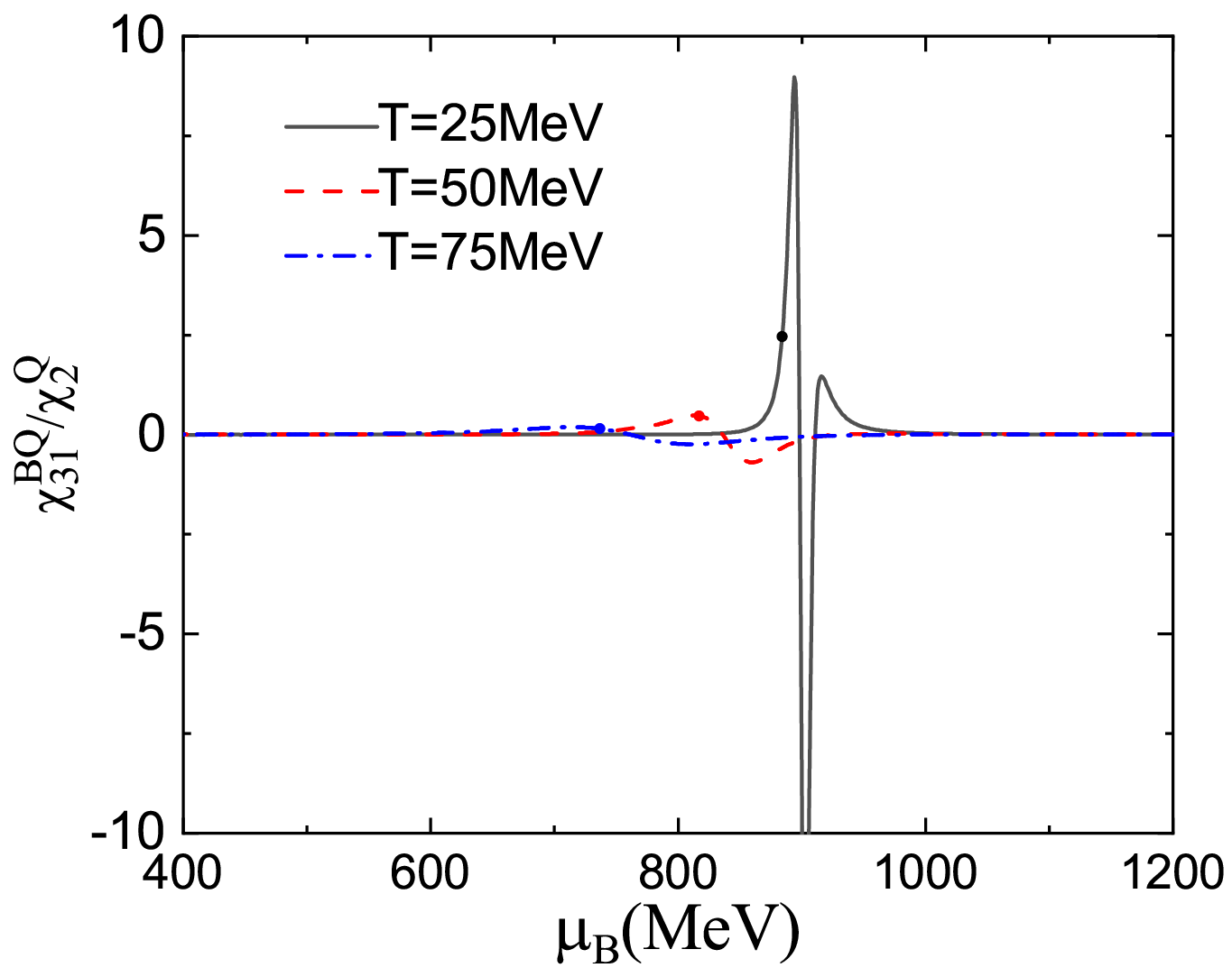}
	\end{center}
	\caption{\label{fig:4} Fourth order correlations between baryon number and electric charge as functions of chemical potential for different temperatures. The solid dots demonstrate the values on the chemical freeze-out line given in Fig.~\ref{fig:1}.
	}
\end{figure}

\begin{figure*}[htbp]
	\begin{center}
		\includegraphics[scale=0.3]{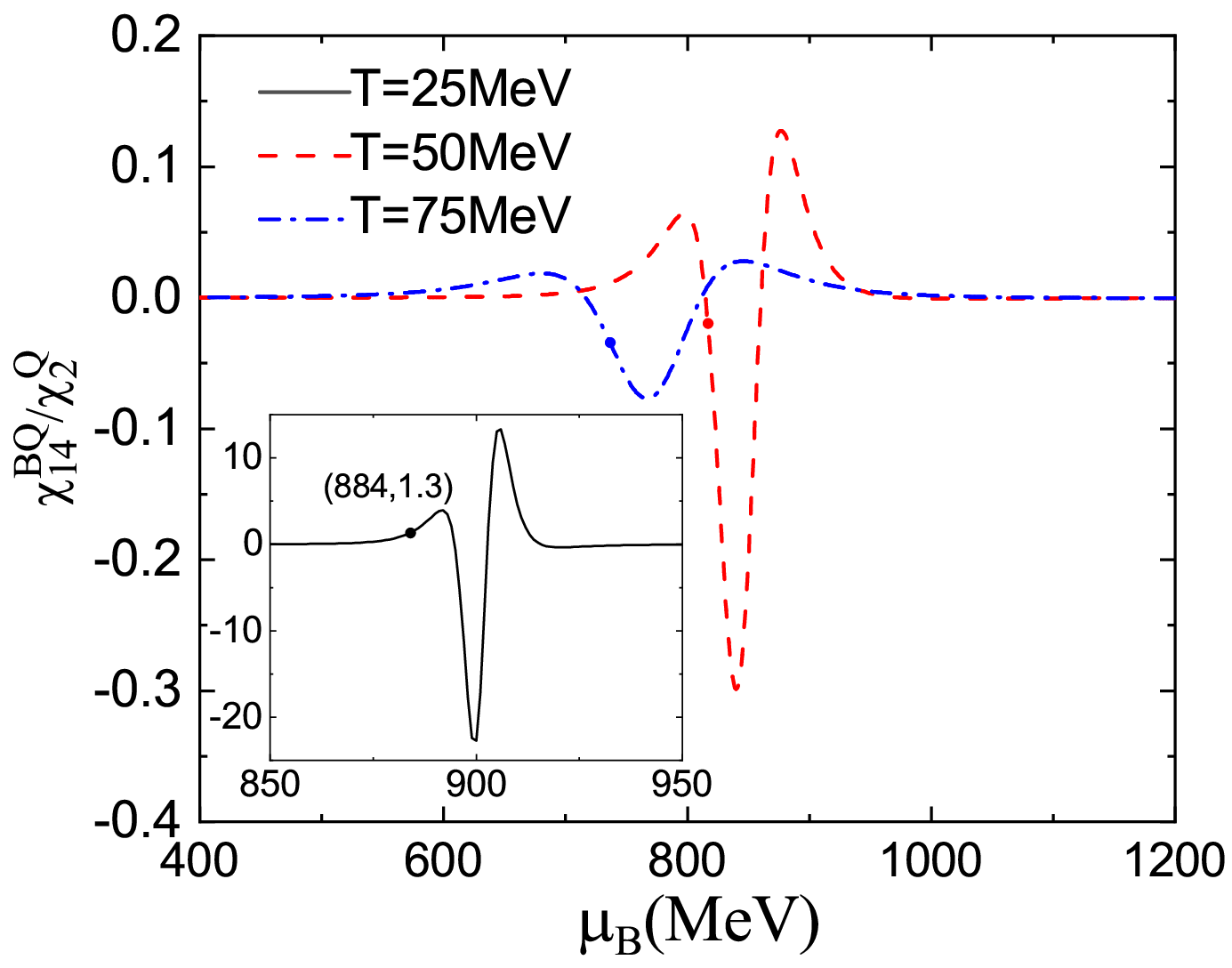}
		\includegraphics[scale=0.3]{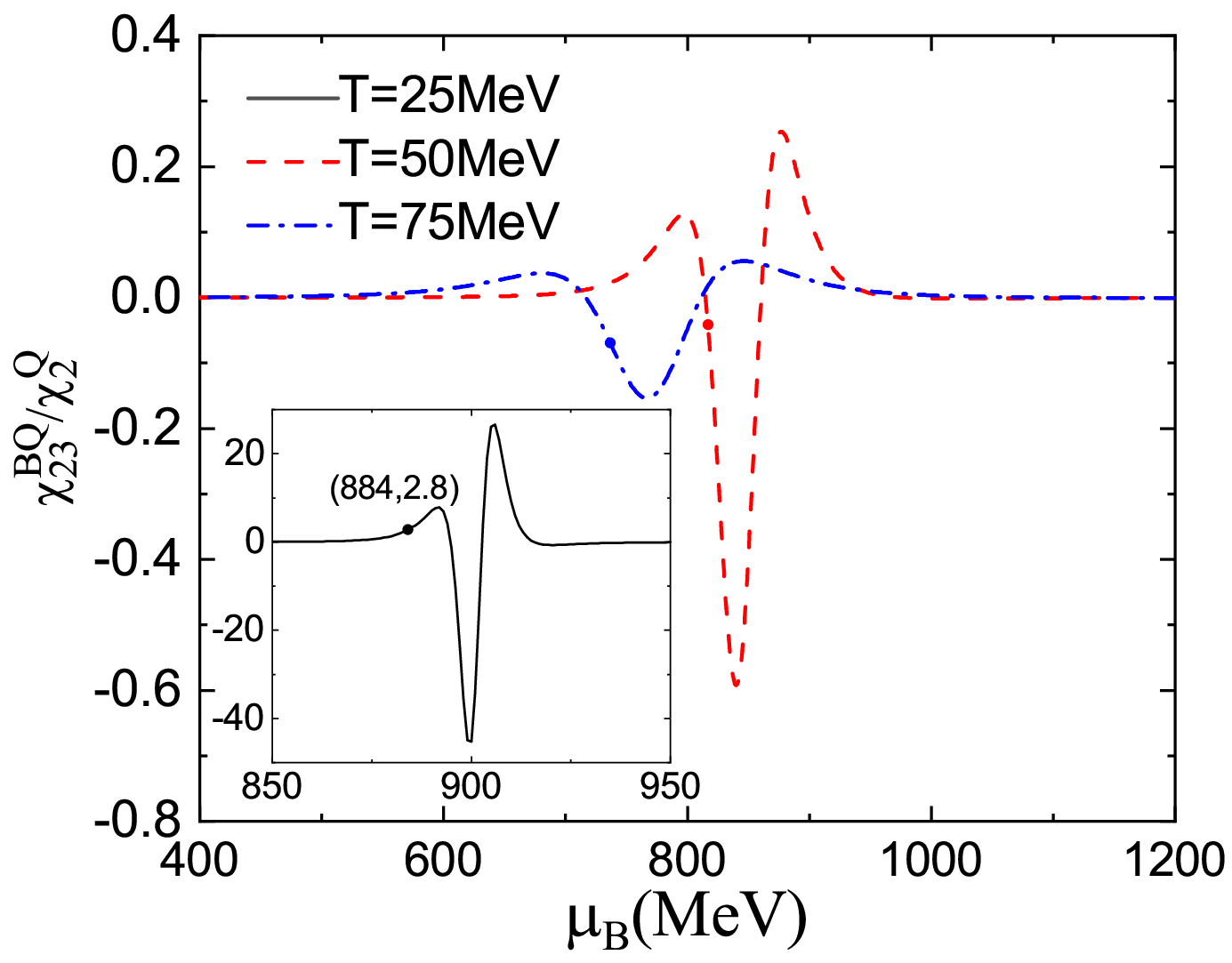}\\
		\includegraphics[scale=0.3]{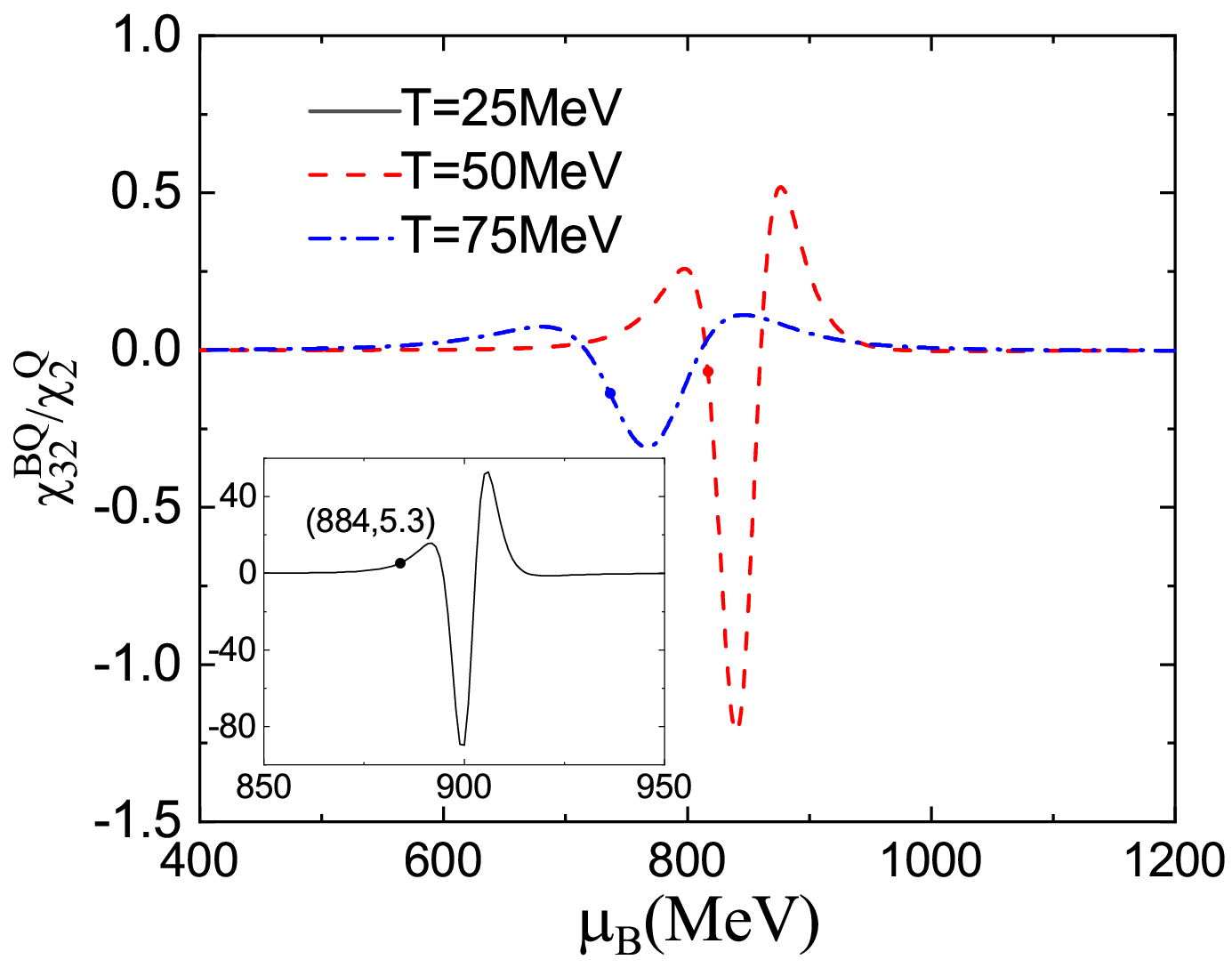}
		\includegraphics[scale=0.3]{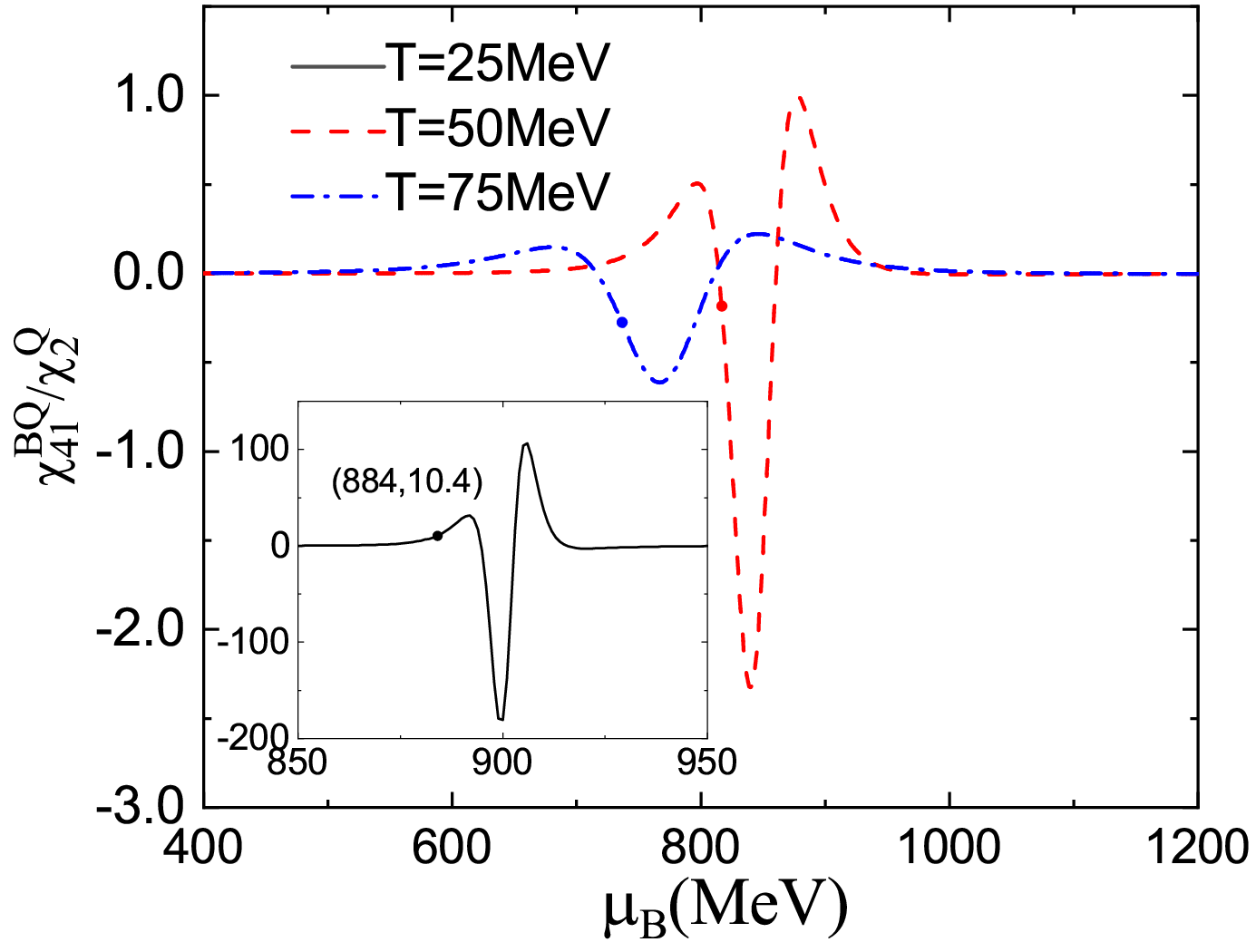}
	\end{center}
	\caption{\label{fig:5} Fifth order correlations between baryon number and electric charge as functions of chemical potential for different temperatures. The solid dots demonstrate the values on the chemical freeze-out line given in Fig.~\ref{fig:1}.
	}
\end{figure*}

In Fig.~\ref{fig:4}, we plot  the fourth order correlations between baryon number and electric charge,  $\chi_{13}^{BQ}/\chi_2^Q$, $\chi_{22}^{BQ}/\chi_2^Q$ and $\chi_{31}^{BQ}/\chi_2^Q$.
Compared with the second and third order correlations in  Fig.~\ref{fig:2} and  Fig.~\ref{fig:3},  Fig.~\ref{fig:4} shows that the rescaled fourth order correlations  by $\chi_2^Q$  are weaker at higher temperature, e.g., $T=75\,$MeV. However, the correlations are much stronger at $T=25\,$MeV, near the critical region of LGPT. Correspondingly, there is evidently a bimodal structure for all the three correlations with the increase of chemical potential at lower temperature. It is also seen that the maximum values of  $\chi_{13}^{BQ}/\chi_2^Q$, $\chi_{22}^{BQ}/\chi_2^Q$ and $\chi_{31}^{BQ}/\chi_2^Q$ increase in turn. Besides, the solid dots demonstrate the value of each correlation at freeze-out increases with the decline of temperature. Moreover,  it is seen that $\chi_{13}^{BQ}/\chi_2^Q < \chi_{22}^{BQ}/\chi_2^Q <\chi_{31}^{BQ}/\chi_2^Q$ at chemical freeze-out for each temperature. It implies that $\chi_{31}^{BQ}/\chi_2^Q$ is most sensitive among the three fourth-order correlations.

Fig.~\ref{fig:5} presents the fifth order correlations between baryon number and electric charge, $\chi_{14}^{BQ}/\chi_2^Q$, $\chi_{23}^{BQ}/\chi_2^Q$  $\chi_{32}^{BQ}/\chi_2^Q$ and $\chi_{41}^{BQ}/\chi_2^Q$ for $T=75, 50, 25\,$MeV.
This figure shows that at $T=75\,$MeV, the values of the three rescaled correlations are all quite small, but they become drastic at $T=25\,$MeV. In combination with the phase diagram in Fig.~\ref{fig:1}, it can been seen that the closer they get to the liquid-gas transition the stronger the high-order correlated fluctuations. Similar to the fourth order correlations, the  rescaled fifth correlations fullfill the relations of $\left|\chi_{14}^{BQ}/\chi_2^Q\right| < \left|\chi_{23}^{BQ}/\chi_2^Q\right|< \left|\chi_{32}^{BQ}/\chi_2^Q\right| < \left|\chi_{41}^{BQ}/\chi_2^Q\right|$ at chemical freeze-out. Moreover, a remarkable result is that all the four fifth-order correlation fluctuations are negative at  chemical freeze-out for $T=75$ and $ 50\,$MeV, but they are positive at $T=25\,$MeV, close to the region of liquid-gas transition. This is a prominent feature in exploring the interaction and phase transition of nuclear matter.

\begin{figure*}[htbp]
	\begin{center}
		\includegraphics[scale=0.3]{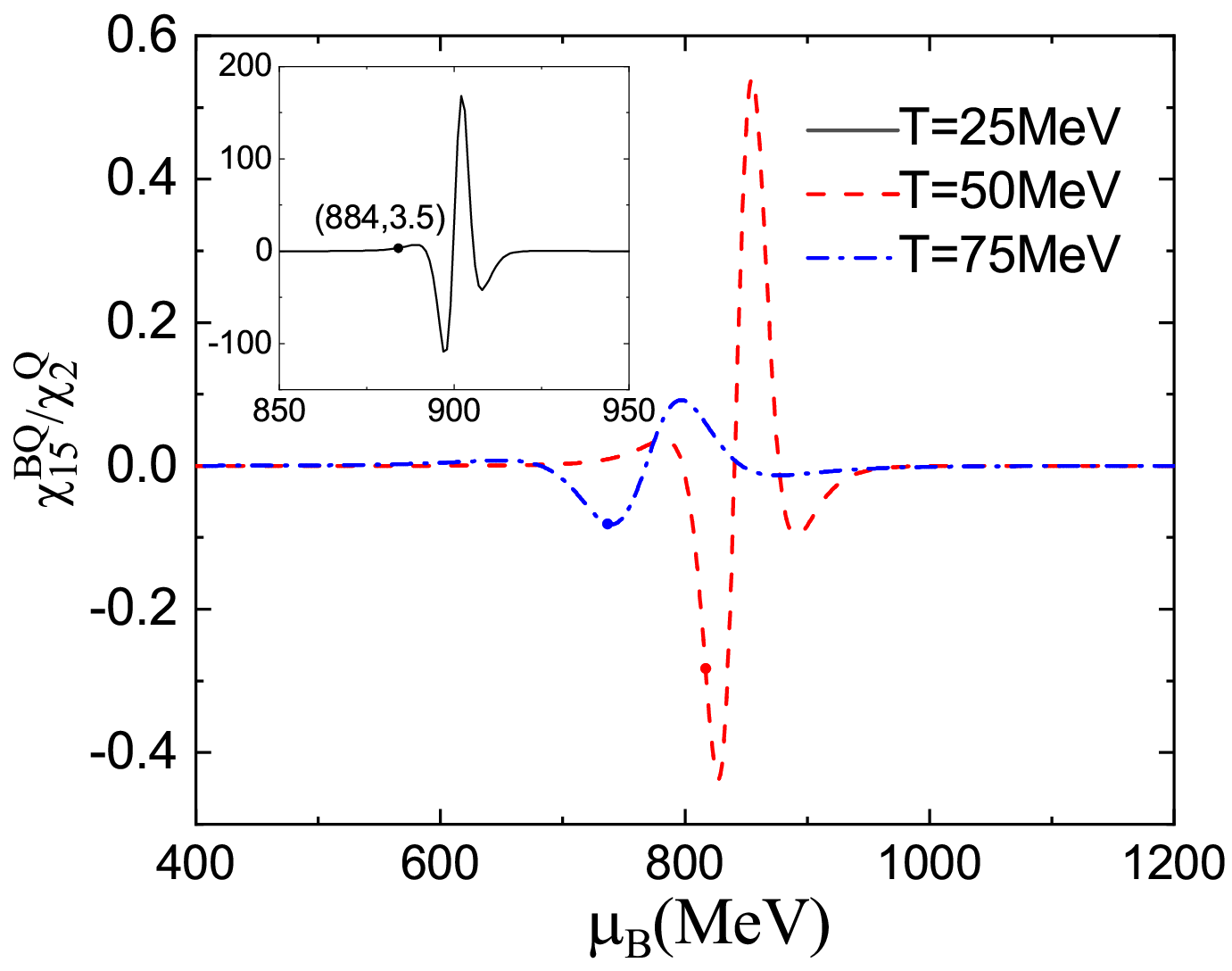}
		\includegraphics[scale=0.3]{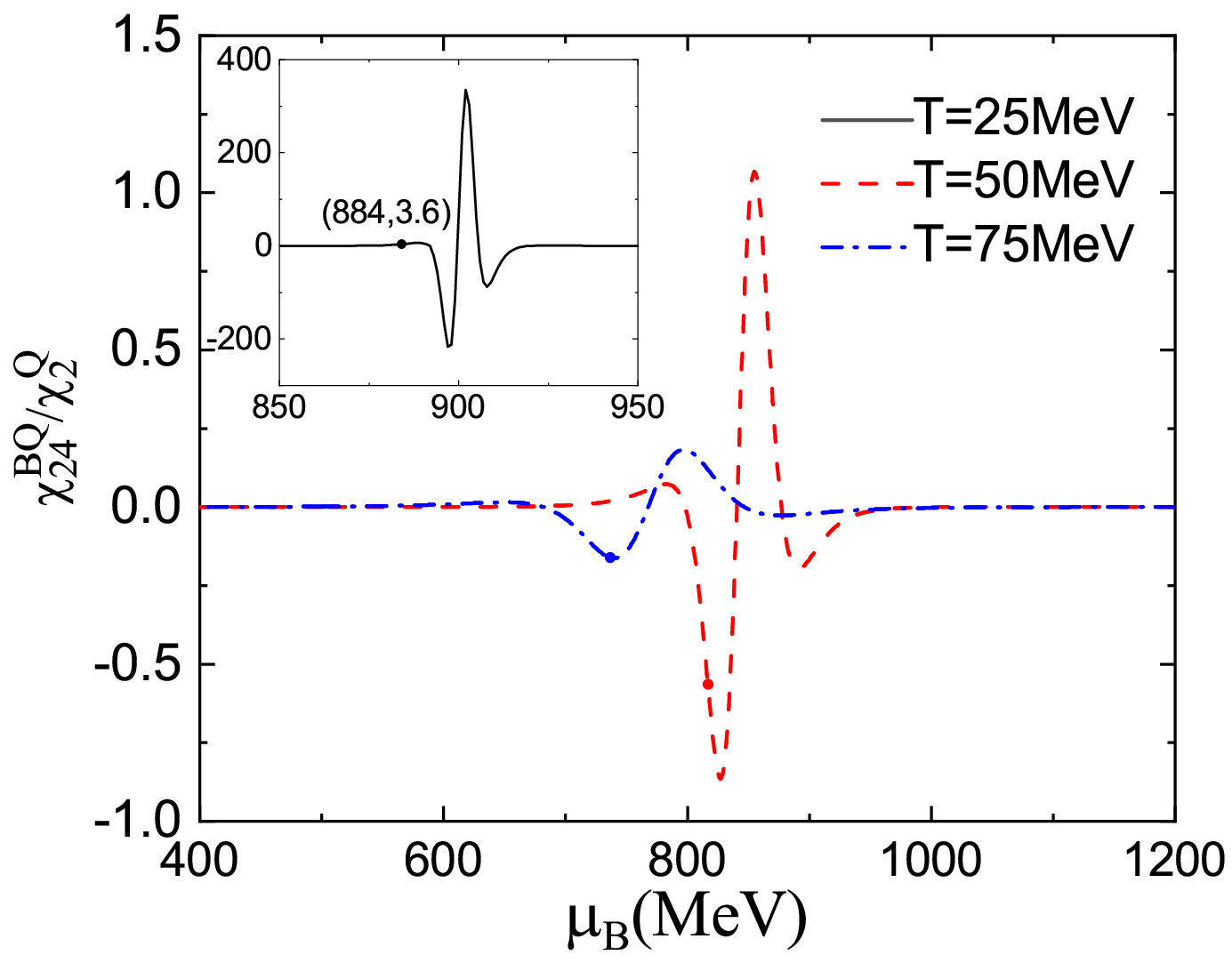}
		\includegraphics[scale=0.3]{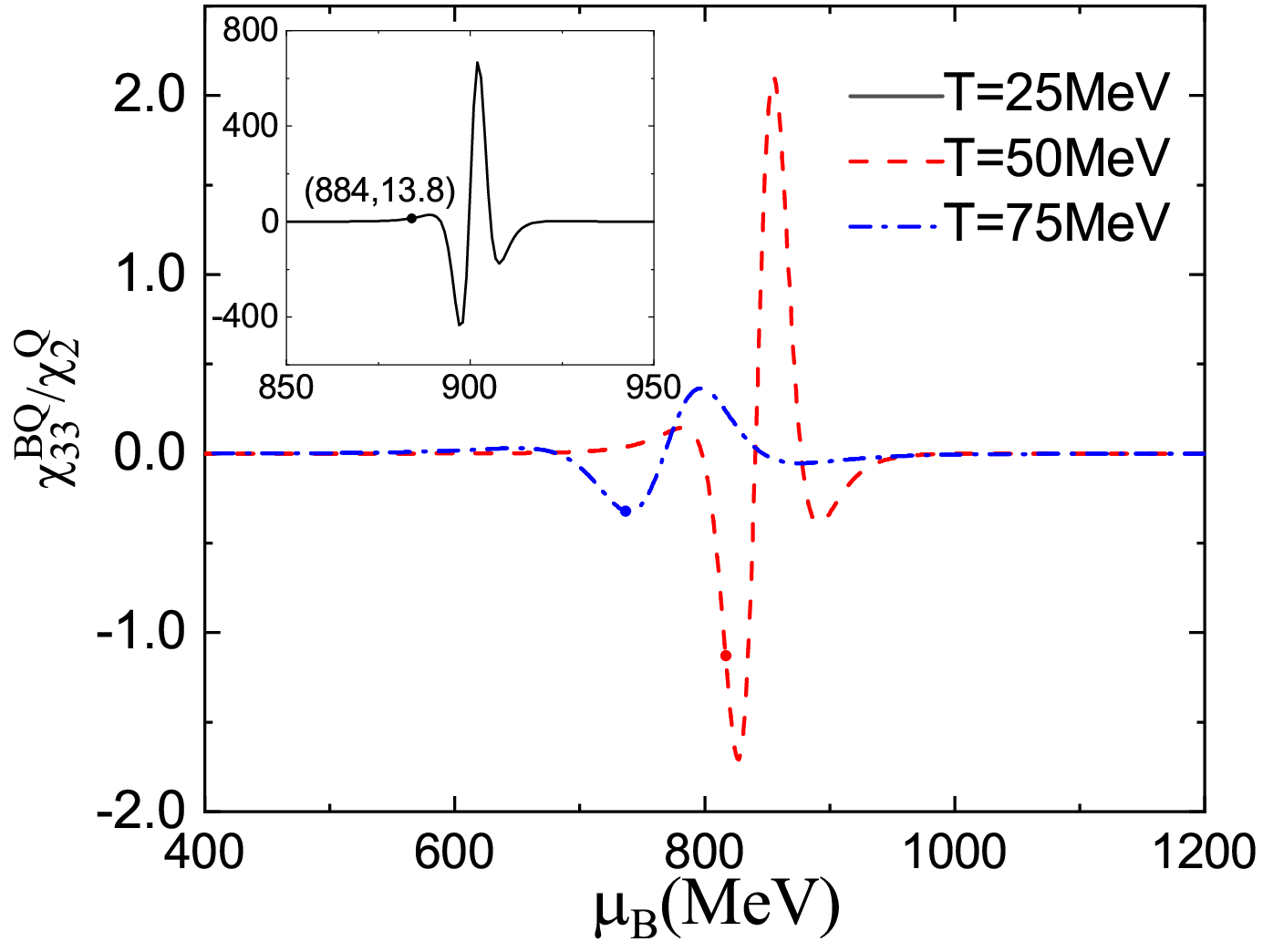}
		\includegraphics[scale=0.3]{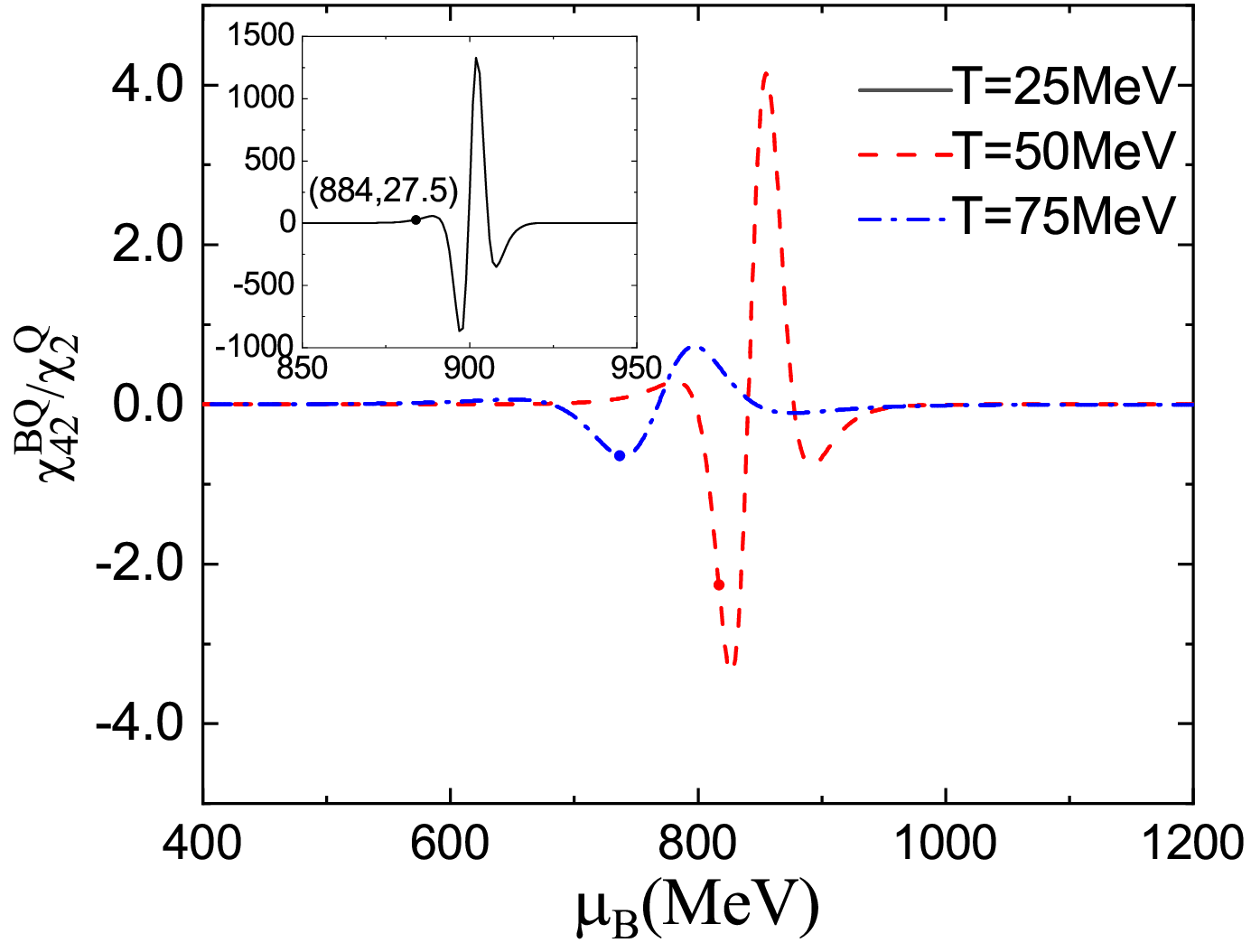} 
		\includegraphics[scale=0.3]{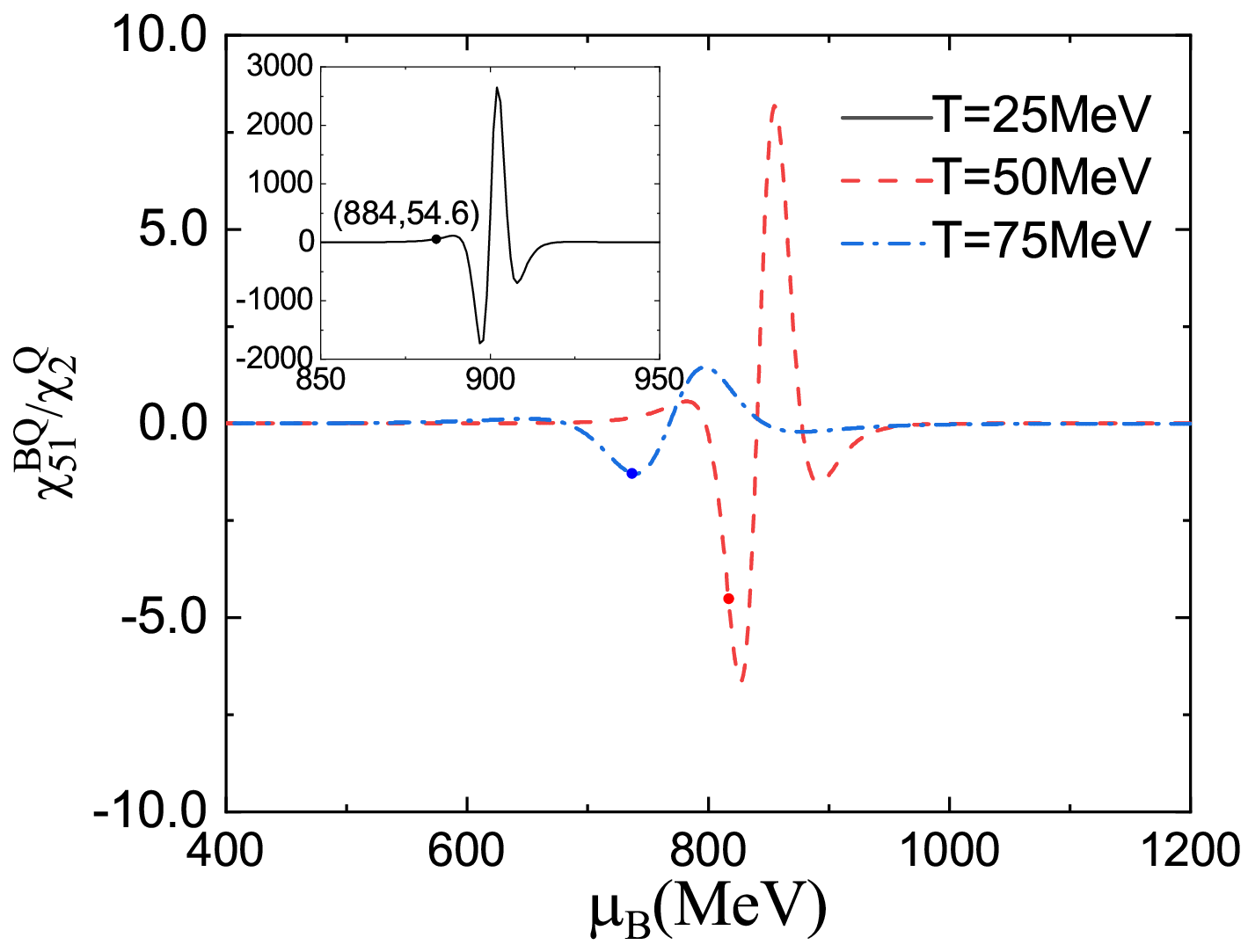}
	\end{center}
	\caption{\label{fig:6} Sixth order correlations between baryon number and electric charge as functions of chemical potential for different temperatures. The solid dots demonstrate the values on the chemical freeze-out line given in Fig.~\ref{fig:1}.
	}
\end{figure*}

Fig.~\ref{fig:6} shows the sixth order correlations of baryon number and electric charge, ie., $\chi_{15}^{BQ}/\chi_2^Q$, $\chi_{24}^{BQ}/\chi_2^Q$  $\chi_{33}^{BQ}/\chi_2^Q$, $\chi_{42}^{BQ}/\chi_2^Q$ and $\chi_{51}^{BQ}/\chi_2^Q$. Each of the sixth order correlations has a double-peak and double-valley structure, although one of the two peaks is not prominent. It is seen that the oscillating behavior intensifies when moving towards the phase transition region from high temperatures to lower ones. Similarly, the intensity of oscillations increases in turn from $\chi_{15}^{BQ}/\chi_2^Q$, $\chi_{24}^{BQ}/\chi_2^Q$  $\chi_{33}^{BQ}/\chi_2^Q$, $\chi_{42}^{BQ}/\chi_2^Q$ to $\chi_{51}^{BQ}/\chi_2^Q$.

For a given order of correlations, the numerical results in Fig.~\ref{fig:2}-\ref{fig:6} show that the signals become stronger when there are more derivatives with respect to baryon chemical potential than that with to electric chemical potential. We also checked the pure baryon number fluctuation, and found it is the most sensitive one at the same order to the LGPT critical end point. The possible reason is that the baryon number fluctuation includes both the proton and neutron's contribution. However, the electric charge fluctuation involves the isospin density, $\rho_N-\rho_P$. The baryon number density is always larger than the isospin density, which is associated with stronger fluctuations when there are more derivatives with respect to baryon chemical potential than that with to electric chemical potential for a given order of correlations. 

Additionally, comparing the results in Fig.~\ref{fig:2}-\ref{fig:6}, we can find that the rescaled higher-order correlations fluctuate more strongly near the phase transition region, while the lower-order correlations at high temperature are relatively larger than  most of the higher-order ones away from the phase transition region. The similar phenomenon exist for the correlations of conserved charges in quark matter~\cite{Fu10}.  According to the fluctuations of net baryon number~\cite{Shao20, Yang2024}, and the correlations between net baryon number and electric charge in this study, it can be seen that the fluctuations and correlations of conserved charges have similar organization structures for nuclear and quark matter. This can be mainly attributed to that the two phase transitions belong to the same universal class and they both describe the interacting matter with temperature and chemical potential dependent fermion masses.

Since the QCD phase transition and nuclear LGPT possibly occur sequentially form high  to low temperature, (even if the LGPT is not triggered) the energy dependent behaviors of fluctuations and correlations can be referenced to look for the phase transition signals of strongly interacting matter. 
Although the latest reported BES II high-precision data at $7.7-39\,$GeV does not show a drastic change of the net baryon number kurtosis, the stronger fluctuation signals possibly appear in heavy-ion experiments with collision energies lower than $7.7$ GeV. Furthermore, in the hadronic interaction dominant evolution with collision energies lower than the threshold of the generation of QGP, the nuclear interaction and phase structure of LGPT will dominate the behavior of fluctuations and correlations of conserved charges. It is worth looking forward to how the fluctuations and correlations change in experiments with the decrease of collision energy.

\section{summary and conclusion}

Fluctuations and correlations of conserved charges are sensitive probes to investigate the phase structure of strongly interacting matter. In this research, we calculated the correlations between net baryon number and electric charge up to sixth order caused by the hadronic interactions in nuclear matter with the non-linear Walecka model, and explored how they relate to nuclear liquid-gas phase transition.

The calculation indicates that the correlations between net baryon number and electric charge  gradually become stronger from the high-temperature region to critical region of nuclear LGPT. In particular, the correlations are drastic at the location where the $\sigma$ field or nucleon mass changes rapidly near the critical region. A similar behavior exists for the chiral crossover phase transition of quark matter. This is mainly attributed to the similar dynamical mass evolution and the same universal class for the chiral phase transition of quark matter and the liquid-gas phase transition of nuclear matter. 

Compared with the lower order correlations, the higher order correlations fluctuate more strongly near the phase transition region, while the rescaled lower order correlations are relatively stronger than most of the higher-order ones away from the phase transition region at high temperature. At the chemical freeze-out for each temperature,  the calculation shows $\chi_{13}^{BQ}/\chi_2^Q < \chi_{22}^{BQ}/\chi_2^Q <\chi_{31}^{BQ}/\chi_2^Q$ for the fourth order correlation,  $|\chi_{14}^{BQ}/\chi_2^Q| < |\chi_{23}^{BQ}/\chi_2^Q |< |\chi_{32}^{BQ}/\chi_2^Q |< |\chi_{41}^{BQ}/\chi_2^Q|$ for the fifth order correlations, and $|\chi_{15}^{BQ}/\chi_2^Q| < |\chi_{24}^{BQ}/\chi_2^Q| < |\chi_{33}^{BQ}/\chi_2^Q| < |\chi_{42}^{BQ}/\chi_2^Q|< |\chi_{51}^{BQ}/\chi_2^Q|$ for the sixth order correlations. 
In particular, the values of fifth and sixth order correlations change from negative to positive when approaching to the critical region of LGPT from  the high-temperature side along the extrapolated chemical freeze-out line.
With the release of more precise data in experiments below 7.7 GeV  in the future, the realistic chemical freeze-out condition can be fitted and the results obtained in this research can be referred to analyze the signals of QCD phase transition and the influence of nuclear liquid-gas phase transition.

\section*{Acknowledgements}
This work is supported by the National Natural Science Foundation of China under
Grant No.~12475145, 11875213, and the Natural Science Basic Research Plan in Shaanxi Province of
China (Program No.~2024JC-YBMS-018).

\end{document}